\documentclass[aps,prl,letterpaper,preprintnumbers,twocolumn,amsmath,amssymb,nofootinbib]{revtex4}

\newcommand{\be}{\begin{equation}}
\newcommand{\ee}{\end{equation}}
\newcommand{\bea}{\begin{eqnarray}}
\newcommand{\eea}{\end{eqnarray}}
\newcommand{\ba}{\begin{array}}
\newcommand{\ea}{\end{array}}
\newcommand{\eref}[1]{(\ref{#1})}

\newcommand{\BC}{{\mathbb C}}

\newcommand{\CP}{{\mathbb C}{\mathbb P}}

\newcommand{\Diff}{{\rm Diff}}
\newcommand{\Gr}{{\rm Gr}}

\newcommand{\tr}[1]{{\rm tr}\, {1}}
\newcommand{\bra}[1]{\langle{1}|}
\newcommand{\ket}[1]{|{1}\rangle}
\newcommand{\ip}[2]{\langle{1}|{2}\rangle}
\newcommand{\vev}[1]{\langle{1}\rangle}

\newcommand{\todo}[1]{{\bf {1}}}
\newcommand{\comment}[1]{}

\def\dd{{\rm d}}

\newcommand{\ft}[2]{{\textstyle\frac{1}{2}}}
\def\fract12{{\textstyle{1\over2}}}
\def\ffract12{\raise .3 em\hbox{$\scriptstyle1$}\kern-.25em/
                \kern-.2em\lower .2 em \hbox{$\scriptstyle2$}}
\def\fractje12{{\scriptstyle{1\over2}}}

\def\part12{{\partial1\over\partial2}}
\def\ex1{e^{\textstyle1}}

\newcommand{\Sec}[1]{\noindent{\em {#1}} ---}

\begin{document}

\preprint{\footnotesize IHES/P/08/25}
\preprint{\footnotesize VPI-IPNAS-08-08}

\title{\Large \bf On the Origin of Time and the Universe}

\author{\bf
Vishnu Jejjala${}^{1}$,
Michael Kavic${}^{2}$,
Djordje Minic${}^{2}$, and
Chia-Hsiung Tze${}^{2}$ \\
${}$ \\}

\affiliation{
${}^1$Institut des Hautes \'Etudes Scientifiques, 35, Route de Chartres, 91440 Bures-sur-Yvette, France \\
${}^2$Institute for Particle, Nuclear and Astronomical Sciences, Department of Physics, Virginia Tech, Blacksburg, VA 24061, U.S.A.
${}$ \\
\footnotesize{\rm vishnu@ihes.fr, kavic@vt.edu, dminic@vt.edu, kahong@vt.edu}
}

\begin{abstract}
We present a novel solution to the low entropy and arrow of time puzzles of the initial state of the Universe.
Our approach derives from the physics of a specific generalization of Matrix theory put forth in earlier work as the basis for a quantum theory of gravity.
The particular dynamical state space of this theory, the infinite dimensional analogue of the Fubini--Study metric over a complex non-linear Grassmannian, has recently been studied by Michor and Mumford.
The geodesic distance between any two points on this space is zero.
Here we show that this mathematical result translates to a description of a hot, zero entropy state and an arrow of time after the Big Bang.
This is modeled as a far from equilibrium, large fluctuation driven, ``freezing by heating'' metastable ordered phase transition of a non-linear dissipative dynamical system.
\end{abstract}

\maketitle

\setcounter{footnote}{0}
\renewcommand{\thefootnote}{\arabic{footnote}}

\Sec{Introduction and summary}
The observed expansion of the Universe together with measurements of the cosmic microwave background radiation vindicate the paradigm of a hot Big Bang.
Standard cosmological models propose an initial spacelike singularity.
Such a state signals the breakdown of spacetime and geometry as effective descriptions of Nature.
Understanding the physics of the singularity and the dynamical evolution of the Universe at the earliest times remains one of the long standing and unrealized ambitions of any putative quantum theory of gravity. % \cite{mtw}.

% The current entropy of the Universe is itself much lower than the maximum possible entropy if all matter were allowed to collapse into black holes~\cite{penrose}.
% While the degrees of freedom of matter came to be in a hot, maximum entropy thermal state, their gravitational counterparts arose in a very special low entropy state.

The initial state of the Universe has a very low entropy.
In fact, from the point of view of the Wheeler--DeWitt equation, the entropy should be zero as the wavefunction of the Universe is unique.
The present entropy of the observed Universe can be estimated by the degrees of freedom associated holographically to the causal horizon:
\be
S\simeq \left(\frac{R_H}{\ell_P}\right)^2 \simeq 10^{123} ~,
\ee
where $R_H$ is the Hubble radius and $\ell_P$ the Planck length.
The number of microstates is then given by Boltzmann's formula $\Omega=e^S\simeq e^{10^{123}}$, and the probability associated with the Big Bang is
\be
P\sim\frac{1}{\Omega}\simeq e^{-10^{123}} ~.
\ee
The Big Bang therefore appears to be an exceptionally special point in phase space, as finely tuned as the cosmological constant~\cite{penrose}.

In this letter, we advance the idea that a low entropy initial state, indeed one with zero entropy, is not only natural but compulsory.
% We address the origin of the Universe in the context of a previously established generalization of quantum theory proposed as a background independent formulation of Matrix theory~\cite{review}.
We address the origin of the Universe in the context of a new approach to quantum gravity
% This formalism is
rooted in a {\em quantum equivalence principle} that renders the state space of a generalized quantum mechanics fully dynamical~\cite{review}.
This indicates that the state space is an infinite dimensional complex non-linear Grassmannian that is a diffeomorphism invariant generalization of $\CP^n$, the complex projective phase space of quantum mechanics~\cite{mt, dj}.

Subsequent to the proposal that this non-linear Grassmannian should play a central role in a theory of quantum gravity, new properties of this space were brought to light that make it uniquely suited for application to the physics of the Big Bang.
According to a remarkable theorem of Michor and Mumford~\cite{mm}, the geodesic distance between any two points on this Grassmannian, as measured by the exact analogue of the Fubini--Study (FS) metric on $\CP^n$, vanishes.
On the strength of this theorem, the everywhere high curvature properties of the metric, and in concert with parallels found in the geometric and topological approach to Hamiltonian dynamics and statistical mechanics of condensed matter systems and in non-equilibrium, dissipative systems, we conclude the following:
(1) That our probabilistic scheme is endowed with a Big Bang event, and because the quantum phase space is comprised of a single microstate this occurs with probability one, implying that $S=0$;
(2) That the Big Bang corresponds to a far from equilibrium collective state, a large fluctuation inducing ``freezing by heating'' metastable phase transition that yields a cosmological arrow of time.

\Sec{Time and M-theory}
Standard quantum mechanics may be cast geometrically as Hamiltonian dynamics over a specific phase space $\CP^n$, the complex projective Hilbert space of pure quantum states~\cite{geoqm}.
$\CP^n$ is a compact, homogeneous, isotropic, and simply connected K\"ahler--Einstein manifold with constant, holomorphic sectional curvature $2/\hbar$.
Notably, being K\"ahler it possesses a triad of compatible structures, any two of which determine the third.
These are a symplectic two-form $\omega$, an unique FS metric $g$, and a complex structure $j$.
All of the key features of quantum mechanics are encoded in this geometric structure.
In particular, the Riemannian metric determines the distance between states on the phase space, and the Schr\"odinger equation is simply the associated geodesic equation for a particle moving on $\CP^n = U(n+1)/(U(n)\times U(1))$ in the presence of an effective external gauge field (namely, the $U(n)\times U(1)$ valued curvature two-form) whose source is the Hamiltonian of a given physical system.
When the configuration space of the theory is the physical space, the FS metric reduces to the spatial metric.
This observation suggests that space, indeed curved spacetime, need not be inputs but may emerge from a suitably extended quantum theory over phase space, generalized both kinematically and dynamically.
To put our results in their proper context, we briefly summarize the pertinent features of the generalized quantum theory.

First, we recall that crucially, Matrix theory is a manifestly second quantized, non-perturbative formulation of M-theory on a fixed spacetime background~\cite{bfss}.
% Bulk spacetime is an emergent feature of the Matrix theory construction.
% The transverse space arises from the parameterization of the moduli space of the Yang--Mills theory on $N$ D$0$-branes.
% The eigenvalues of the matrices denote the classical positions of the D-branes.
% The interactions of the branes via open strings are specified by off-diagonal entries in the matrices.
% Since the matrices are a description of the target space, the spacetime geometry becomes non-commutative once interactions turn on.
% As D$0$-branes are themselves graviton bound states, the gravitational interaction and thus the geometry of spacetime are contained in the open string dynamics, {\em viz.}\ the quantum fluctuations of the matrix degrees of freedom.
While physical space emerges as a moduli space of the supersymmetric matrix quantum mechanics, time still appears as in any other canonical quantum theory.

Time is not an observable in quantum mechanics:
there is no ``clock'' operator.
Moreover, as we demand diffeomorphism invariance in a theory of gravitation, time and spatial position are simply labels, and when the metric is allowed to fluctuate, classical notions, such as spacelike separation of points, cease to have operational meaning.
To construct a background independent
% \footnote{
% By {\em background independence} we simply mean that no {\em a priori} choice of asymptopia is made in the configuration space.}
formulation of Matrix theory, it becomes necessary to relax the rigidity of the underlying quantum theory.

% We do so by starting from a larger phase space, larger in the sense of symmetries as well as making it dynamical.
% In the spirit of the correspondence principle, we expect the resulting theory to give rise, in some appropriate limit, to an emergent Lorentzian spacetime along with the Hilbert space based M-Theory quantum mechanics.

The extension of geometric quantum mechanics via a quantum equivalence principle yields the following~\cite{review, mt, jm}.
At the basic level, there are only dynamical correlations between quantum events.
The phase space must have a symplectic structure, namely a symplectic two-form, and be the base space of a $U(1)$ bundle; and it must be diffeomorphism invariant.
We demand a three-way interlocking of the Riemannian, the symplectic, and the non-integrable almost complex structures.
% Straying away from the integrable complex structure of $\CP^n$ is in fact the {\it only} change made to the structure of quantum mechanical state space.
% Yet just such a weakening move, physically motivated by going from a global to a local time, a notion more in line with general relativity, requires significant changes, kinematical and dynamical.
% First, one must go beyond the linear unitary groups $SU(n)$ to the non-linear infinite dimensional diffeomorphism groups.
% Secondly, a phase space is singled out, one admitting a manifold of metrics:
% it is the non-linear Grassmannian
In departing from the integrable complex structure of $\CP^n$, the quantum mechanical phase space becomes the non-linear Grassmannian,
$\Gr(\BC^{n+1}) = \Diff(\BC^{n+1})/\Diff(\BC^{n+1}, \BC^n\times \{0\})$,
with $n\to\infty$, a complex projective, strictly almost K\"ahler manifold.
Moreover, diffeomorphism invariance implies that not just the metric but also the almost complex structure and hence the symplectic structure be fully dynamical.
Consequently, with the coadjoint orbit nature of $\Gr(\BC^{n+1})$, the equations of motion of this general theory are the Einstein--Yang--Mills equations:
\be
\label{BIQM1}
{\cal{R}}_{ab} - \frac{1}{2} {\cal{G}}_{ab} {\cal{R}}  - \lambda {\cal{G}}_{ab}= {\cal{T}}_{ab} (H, F_{ab}) ~,
\ee
with ${\cal{T}}_{ab}$ as determined by ${\cal{F}}_{ab}$, the holonomic Yang--Mills field strength, the Hamiltonian (``charge'') $H$, and a ``cosmological'' term $\lambda$.
Furthermore,
\be
\label{BIQM2}
\nabla_a {\cal{F}}^{ab} = \frac{1}{2M_P} H u^b ~,
\ee
where $u^b$ are the velocities, $M_P$ is the Planck energy, and $H$ the Matrix theory Hamiltonian~\cite{bfss}.
These coupled equations imply via the Bianchi identity a conserved energy-momentum tensor: $\nabla_a {\cal{T}}^{ab} =0$.
Just as the geodesic equation for a non-Abelian charged particle is contained in the classical Einstein--Yang--Mills equations, so is the corresponding geometric, covariant Schr\"odinger equation.
It is here genuinely non-linear and cannot be, as in quantum mechanics, linearized by lifting to a flat Hilbert space.
% As for time, we must first solve for the metric and then find out what the associated time function is just as in general relativity.
The above set of equations defines the physical system (here the model Universe) and identifies the correct variables including time.

\Sec{Geometry of $\Gr(\BC^{n+1})$}
As the space $\Gr(\BC^{n+1})$ is the central focus of this letter, and for comparison to $\CP^n$, we list its main features.
It is a compact, homogeneous but non-symmetric, multiply-connected, infinite dimensional complex Riemannian space.
It is a projective strictly almost K\"ahler manifold, a coadjoint orbit, hence a symplectic coset space of the volume preserving diffeomorphism group~\cite{hv}.
It is also the base manifold of a circle bundle over $\Gr(\BC^{n+1})$, where the $U(1)$ holonomy provides a Berry phase.

Crucial for our purposes, non-linear Grassmannians are {\em Fr\'echet spaces}.
As generalizations of Banach and Hilbert spaces, Fr\'echet spaces are locally convex and complete topological vector spaces.
(Typical examples are spaces of infinitely differentiable functions encountered in functional analysis.)
Defined either through a translationally invariant metric or by a countable family of semi-norms, the lack of a true norm makes their topological structures more complicated.
The metric, not the norm, defines the topology.
% CHIA: Since topology and metric are tied here, it would be incorrect to say that something is metrical and not topological. Unsurprisingly, we deal with some strange creatures here !
Moreover, there is generally no natural notion of distance between two points so that many different metrics may induce the same topology.
In sharp contrast to $\CP^n$, the allowed metrical structures are much richer and more elastic, thereby allowing novel probabilistic and dynamical applications.
Thus $\Gr(\BC^{n+1})$ has in principle an infinite number of metrics, a subset of which form the solution set to the Einstein--Yang--Mills plus Matrix model equations we associate with the space.
For example, in~\cite{mm}, an infinite one-parameter family of non-zero geodesic distance metrics are found.

% \Sec{The Michor--Mumford vanishing theorem}
Since $\Gr(\BC^{n+1})$ is the diffeomorphism invariant counterpart of $\CP^n$, the simplest and most natural topological metric to consider is the analogue of the FS metric.
This weak metric was analyzed by Michor and Mumford~\cite{mm}, who obtained the striking result, henceforth called their {\em vanishing theorem}.
% While it applies to all non-linear Grassmannians, for one of them, $\Gr(\BC^{n+1})$, which has a symplectic two-form and is a candidate space for Hamiltonian dynamics, this theorem states that
The theorem states that the generalized FS metric induces on $\Gr(\BC^{n+1})$ a vanishing geodesic distance.
Such a paradoxical phenomenon is due to the curvatures being unbounded and positive in certain directions causing the space to curl up so tightly on itself that the infinitum of path lengths between any two points collapses to zero.

\Sec{A Universe of zero size}
The crucial point of this work is to take seriously this most unusual mathematical property of $\Gr(\BC^{n+1})$ and to interpret it in physical terms.
Taking this as the space of states out of which spacetime emerges, we see that the vanishing theorem naturally describes an initial state in which the Universe exists at single point, the cosmological singularity.

Moreover, viewed through this lens, a statistical notion of time may apply close to the cosmological singularity.
We observe that in both the standard geometric quantum mechanics and its extension, the Riemannian structure encodes the statistical structure of the theory.
The geodesic distance is a measure of change in the system, for example through Hamiltonian time evolution.
By way of the FS metric and the energy dispersion $\Delta E\ $, the infinitesimal distance in phase space is
\be
\dd s = \frac{2}{\hbar}\, \Delta E\, \dd t ~.
\label{distance}
\ee
Through this relation, time reveals its statistical, quantum nature.
It also suggests that dynamics in time relate to the behavior of the metric on the configuration space.

%  Chia here : Hi guys, just as that for the expansion of the metric g that I deleted (!), to save precious space this trivial Aharonov-Anandan relation should be made part of the text. I don't quite know how to latex this in. So please do so ! Thanks !

% Near the initial singularity spacetime becomes statistical and highly curved.
% Such an initial cosmological phase is the primary testing ground for any quantum theory of gravity.
% It is there that our scheme is expected to tackle unanswered questions about the very early universe.

As Wootters~\cite{woot} showed, what the geodesic distance $\dd s$ on $\CP^n$ measures is the optimal distinguishability of nearby pure states:
if the states are hard to resolve experimentally, then they are close to each other in the metrical sense.
Statistical distance is therefore completely fixed by the size of fluctuations.
A telling measure of the uncertainty between two neighboring states or points in the state space is given by computing the volume of a spherical ball $B$ of radius $r$ as $r\to0$ around a point $p$ of a $d$-dimensional manifold ${\cal M}$.
This is given by
\be
\frac{\textrm{Vol}( B_p(r) )}{\textrm{Vol}( B_e(1) )} = r^d \left( 1 - \frac{R(p)}{6(d+1)} r^2  + o(r^2) \right) ~,
\label{balls}
\ee
where the left hand side is normalized by $\textrm{Vol}( B_e(1) )$, the volume of the $d$-dimensional unit sphere.
$R(p)$, the scalar curvature of ${\cal M}$ at $p$, can be interpreted as the average statistical uncertainty of any point $p$ in the state space~\cite{petz}.
As $2/\hbar$ is the sectional curvature of $\CP^n$, $\hbar$ can be seen as the mean measure of quantum fluctuations.
Eq.~\eref{balls} indicates that, depending on the signs and values of the curvature, the metric distance gets enlarged or shortened and may even vanish.

The vanishing geodesic distance under the weak FS metric on $\Gr(\BC^{n+1})$ is completely an effect of extremely high curvatures~\cite{mm}.
Because the space is extremely folded onto itself, any two points are indistinguishable ({\em i.e.}\ the distance between them is zero).
This is an exceptional locus in the Fr\'echet space of all metrics on $\Gr(\BC^{n+1})$.
% In the words of Michor and Mumford, it is a first actual example of this purely infinite dimensional phenomenon, a most counter-intuitive one which is not available in the $\CP^n$ of standard quantum theory.
This is a purely infinite dimensional phenomenon, and one that does not occur with the $\CP^n$ of the canonical quantum theory.

%Could such a metric be a solution to our generalized Einstein--Yang--Mills theory over $\Gr(\BC^{n+1})$?
%While an affirmative answer would be enticing, to our knowledge, no proof is technically feasible, as is generally the case.
% in particular with such an infinite dimensional non-linear, complex valued, dynamical system.
%Yet we shall argue on compelling physical grounds in what sense this relates closely to the dynamics of our model.

\Sec{The low entropy puzzle}
From the foregoing discussion, the low entropy problem tied to the initial conditions of the Universe is naturally resolved.
In the language of statistical geometry and quantum distinguishability, the generalized FS metric having vanishing geodesic distance between any two of its points means that none of the states of our non-linear Grassmannian phase space can be differentiated from each other.
Due to the large fluctuations in curvatures everywhere, the whole phase space is comprised of a single, {\em unique} microstate.
Since the state space is the model for quantum cosmology, if its metric is the weak Michor--Mumford FS metric, the Universe is in a fixed configuration with probability one.
As we shall see, this is a non-equilibrium setting, but we may nevertheless infer via Boltzmann's formula~\cite{gs} that the entropy of the Universe is identically zero.

\Sec{The Big Bang as the ultimate traffic jam}
What could the physics behind a low (zero) entropy, yet high temperature state of the Big Bang be?
We suggest that the paradoxical zero distance, everywhere high curvature property of $\Gr(\BC^{n+1})$ with the FS metric finds an equally paradoxical physical realization in the context of our model.
This is to be found in a class of far from equilibrium collective phase transitions, the so called ``freezing by heating'' transitions.
From many studies~\cite{pettini} it has been established that high curvatures in the phase or configuration manifold of a physical system precisely reflect large fluctuations of the relevant physical observables at a phase transition point.
This correspondence means equating the high curvatures of the FS metric on $\Gr(\BC^{n+1})$ with large fluctuations in our system at a phase transition.
The vanishing geodesic distance can be interpreted as the signature, or order parameter, of a strong fluctuation (or ``heat'') induced zero entropy and hence highly ordered state.

While from an equilibrium physics perspective such a state seems nonsensical, it occurs in certain far from equilibrium environments.
Specifically, we point to a representative continuum model~\cite{helbing, zia} where such an unexpected state was first discovered.
Here, one has a system of particles interacting, not only through frictional forces and short range repulsive forces, but also and most importantly via strong driving fluctuations ({\em e.g.}, noise, heat, etc.).
As the amplitude of the fluctuations ({\em e.g.}, temperature) goes from weak to strong to extremely strong and as its total energy increases, such a system shows a thermodynamically counterintuitive evolution from a fluid to a solid and then to a gas.
At and beyond the onset of strong fluctuations, it first goes to a highly ordered, low entropy, indeed a crystalline state, which is a phase transition like-state if both particle number and fluctuations are sufficiently large.
This collective state, being energetically metastable then goes into a third disordered, higher entropy gaseous state under extremely strong fluctuations.

While our model's dynamics are mathematically far more intricate than the above models for phenomena such as traffic jams and the flocking of birds, it does have the requisite combination of the proper kind of forces to achieve these ``freezing by heating'' transitions.
%In fact by mere inspection, the following key qualitative features can be read off.
The system being considered is far from equilibrium with low entropy, high temperature, and negative specific heat.
In addition we have non-linear, attractive, and repulsive Yang--Mills forces, short range repulsive forces of D$0$-branes in the Matrix theory, repulsive forces from a positive ``cosmological'' term, and most importantly large gravitational fluctuations induced by the large curvatures.
Moreover it is known that geometric quantum mechanics can be seen as a classical Hamiltonian system, one with a K\"ahler phase space.
Its complete integrability in the classical sense~\cite{block} derives from this K\"ahler property which returns hermiticity of all observables in their operatorial representations.
The extended quantum theory is similarly viewed in terms of classical non-linear field and particle dynamics over a strictly almost complex phase space.
This last property implies that corresponding operators are non-hermitian, and hence this is a dissipative system~\cite{rajeev}.
Moreover, classical Einstein--Yang--Mills systems are non-integrable and chaotic~\cite{barrow}.

\Sec{Time's arrow}
From the relation between geodesic distance and time, we also have the emergence of a cosmological arrow of time.
While the system has entropy $S=0$, the very high curvatures in $\Gr(\BC^{n+1})$ signal a non-equilibrium condition of dynamical instability.
Because of its non-linear dissipative and chaotic dynamics, our system will flow toward differentiation, which thereby yields, through entropy production, distinguishable states in the state space.
This instability is further evidenced by the above mentioned existence of a whole family of non-zero geodesic distance metrics, of which the zero entropy metric is a special case~\cite{mm}.
The dynamical evolution according to the second law is toward some higher entropy but stable state.
% It is at some plateau of
During this evolution, spacetime, and canonical quantum mechanics emerge.

Furthermore, the model we have presented is a generalized quantum dissipative system, {\em i.e.}\ one with frictional forces at work.
%Even if the Michor--Mumford FS metric turns out not to be an exact solution of our system of equations, its compelling
%interpretation as a phase transition indicates that its form may be universal in that it is not sensitive to details of
%the underlying dynamics of the system.
% Non-linearity and dissipation suggest the picture of our
%Universe as a possibly self-organizing, open system the way many systems are in nature.
% It is embedded in an yet larger, even more mysterious environment, one which brings it
%into existence through a low entropy hot Big Bang.
% This last bit is too speculative for me. -V.
Because the fluctuations of linear quantum mechanics and its associated equilibrium statistical mechanics are incapable of driving a system such as our Universe to a hot yet low entropy state and of generating a cosmological arrow of time~\cite{penrose2}, a non-linear, non-equilibrium, strong fluctuation driven quantum theory such as the one presented here becomes necessary.
Time irreversibility is of course a hallmark of non-equilibrium systems;
%, and
%with negative specific heat it thus this
this cosmological model naturally produces both an arrow and an origin of time.
Moreover, in this approach the relationship of canonical quantum theory and equilibrium statistical mechanics is extended to an analogy of generalized quantum theory and non-equilibrium statistical mechanics.
%Emboldened by the solution reported here,
%We are exploring

An interesting avenue of further investigation is the possible extrapolation of the results concerning $\Gr(\BC^{n+1})$ to the study of black hole singularities.

${}$ \\

\medskip
\Sec{Acknowledgments}
{\em This paper is dedicated to John Archibald Wheeler, who inspired us.}
We gratefully acknowledge stimulating discussions with Joe Polchinski, Beate Schmittmann, Tatsu Takeuchi, and Royce Zia.
CHT specifically wishes to thank Cornelia Vizman for first alerting him to the key paper by Michor and Mumford on vanishing geodesic distance and Mathieu Molitor for many mathematical clarifications on non-linear Grassmannians.
VJ acknowledges the ICTP, Trieste for providing a stimulating working environment and LPTHE, Jussieu for hospitality.
MK thanks Bing Feng, Greg Stock, and Roger A.\ Wendell for thoughtful insights.
%and perceptive insights.
DM \& MK are supported in part by the U.S. Department of Energy under contract DE-FG05-92ER40677, task A.

\end{document}